\documentclass[journal]{IEEEtran}
\usepackage{cite}
\usepackage{amsmath,amssymb,amsfonts}
\usepackage{algorithmic}
\usepackage{graphicx}
\usepackage{textcomp}

\usepackage{comment}
\usepackage{amsmath,amssymb} % define this before the line numbering.
\usepackage{color}
\usepackage{subcaption}
\usepackage{blindtext}
\usepackage{paralist, tabularx}

\usepackage{epsfig}
\usepackage{siunitx}
\usepackage{mathtools}
\usepackage{booktabs}       % professional-quality tables
\usepackage{multirow}
\usepackage{bbm}
\usepackage{caption}
\usepackage[table,xcdraw]{xcolor}
\usepackage{pifont}
\usepackage{mwe}
\usepackage{cuted}
\usepackage{lipsum}
\usepackage{array}
\newcommand{\PreserveBackslash}[1]{\let\temp=\\#1\let\\=\temp}
\newcolumntype{C}[1]{>{\PreserveBackslash\centering}p{#1}}
\newcolumntype{R}[1]{>{\PreserveBackslash\raggedleft}p{#1}}
\newcolumntype{L}[1]{>{\PreserveBackslash\raggedright}p{#1}}

\newcommand{\fref}[1]{Figure~\ref{#1}}
\newcommand{\sref}[1]{Section~\ref{#1}}
\newcommand{\tref}[1]{Table~\ref{#1}}

\newcommand{\xmark}{\ding{55}}
\newcommand{\cmark}{\ding{51}}
\usepackage{floatrow}
\newfloatcommand{capbtabbox}{table}[][\FBwidth]
\usepackage[switch]{lineno}

\begin{document}
% \linenumbers
% \lipsum[1-20]

%
% paper title
% Titles are generally capitalized except for words such as a, an, and, as,
% at, but, by, for, in, nor, of, on, or, the, to and up, which are usually
% not capitalized unless they are the first or last word of the title.
% Linebreaks \\ can be used within to get better formatting as desired.
% Do not put math or special symbols in the title.
\title{Indirect Domain Shift for Single Image Dehazing}
%
%
% author names and IEEE memberships
% note positions of commas and nonbreaking spaces ( ~ ) LaTeX will not break
% a structure at a ~ so this keeps an author's name from being broken across
% two lines.
% use \thanks{} to gain access to the first footnote area
% a separate \thanks must be used for each paragraph as LaTeX2e's \thanks
% was not built to handle multiple paragraphs
%

\author{Huan Liu,~\IEEEmembership{Student Member,~IEEE,}
        and~Jun Chen,~\IEEEmembership{Senior Member,~IEEE}% <-this % stops a space
\thanks{Huan Liu  and Jun Chen are with the Department
of Electrical and Computer Engineering, McMaster University, Hamilton,
ON, L8S 2K1 CA e-mail: liuh127@mcmaster.ca.}% <-this % stops a space% <-this % stops a space
\thanks{Manuscript received April 19, 2005; revised August 26, 2015.}}

\maketitle

% As a general rule, do not put math, special symbols or citations
% in the abstract or keywords.
\begin{abstract}
Despite their remarkable expressibility, convolution neural networks (CNNs) still fall short of delivering satisfactory results on single image dehazing,  especially in terms of faithful recovery of fine texture details. In this paper, we argue that the inadequacy of conventional CNN-based dehazing methods can be attributed to the fact that the domain of hazy images is too far away from that of clear images, rendering it difficult to train a CNN for learning direct domain shift through an end-to-end manner and recovering texture details simultaneously.  To address this issue, we propose to add explicit constraints inside a deep CNN model to guide the restoration process.
In contrast to direct learning, the proposed mechanism shifts and narrows the candidate region for the estimation output via multiple confident neighborhoods. Therefore, it is capable of consolidating the expressibility of different architectures, resulting in a more accurate indirect domain shift (IDS) from the hazy images to that of clear images.
We also propose two different training schemes, including hard IDS and soft IDS, which further reveal the effectiveness of the proposed method.
Our extensive experimental results indicate that the dehazing method based on this mechanism dramatically outperforms the state-of-the-arts.
\end{abstract}

% Note that keywords are not normally used for peerreview papers.

% For peer review papers, you can put extra information on the cover
% page as needed:
% \ifCLASSOPTIONpeerreview
% \begin{center} \bfseries EDICS Category: 3-BBND \end{center}
% \fi
%
% For peerreview papers, this IEEEtran command inserts a page break and
% creates the second title. It will be ignored for other modes.
\IEEEpeerreviewmaketitle

\section{Introduction}

	Deep convolutional neural networks (CNNs) have been tremendously successful in many high-level computer vision tasks, e.g. image recognition \cite{Krizhevsky:2012wl,He:2016ib} and object detection \cite{Girshick:2015id,Ren:2017kt}.
	Although recent works have shown that it is also possible to learn an end-to-end CNN model for low-level vision tasks, e.g. image dehazing ~\cite{Cai:2016ge,Johnson:2016hp}, the resulting performance is still not completely satisfactory. 	For high-level vision tasks, it suffices to extract specific features  and simply express them as very low dimensional vectors ~\cite{Krizhevsky:2012wl}, which results in a relatively simple mapping.
	In contrast, low-level vision tasks require both global understanding of image content and local inference of texture details; as such, the associated mappings are more complicated.

	One possible explanation for performance discrepancies on high-level and low-level vision tasks is as follows. For high-level vision tasks such as image recognition, a slight perturbation of the output tends to be inconsequential since the perturbed output is likely to get converted to the same one-hot vector and consequently the classification label remains unaffected. However, for low-level vision tasks such as image dehazing, any perturbation can potentially manifest in the final result, jeopardizing the image quality.
	From this point of view, despite the fact that a deep CNN can in principle approximate any function, it is still difficult to train an accurate mapping that lifts the input to the target domain in one shot, since the loss function is typically very close to zero  in the neighborhood of the target image ~\cite{Lin:2017de}. We argue that a different mechanism for domain shift is needed for image dehazing, which requires both memory and understanding of image contents.

\begin{figure*}[!t]
	\begin{center}
		\includegraphics[width=0.9\linewidth]{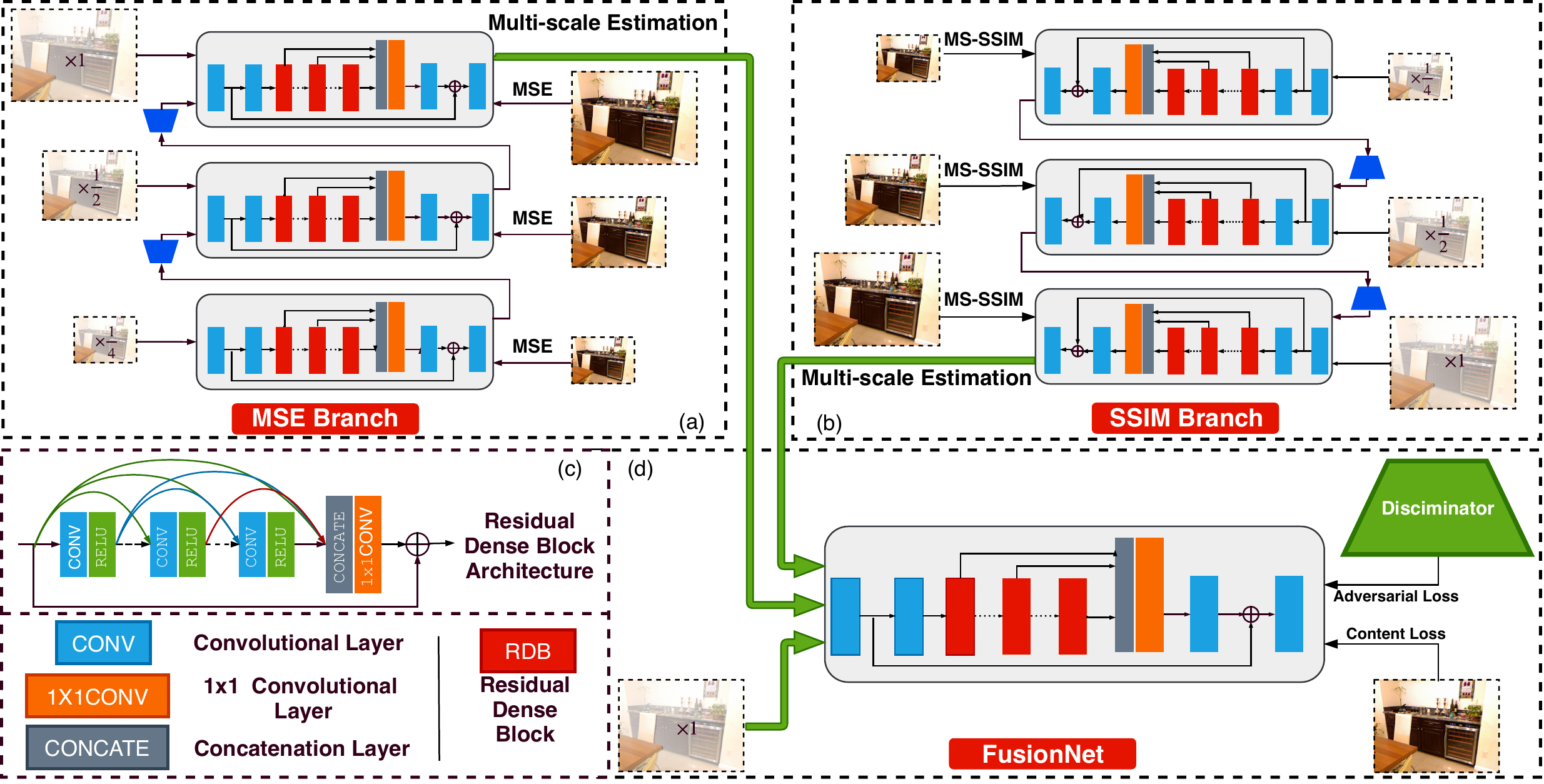}
	\end{center}
	\caption{One example of the proposed IDS network. (a) and (b) are the multi-scale estimation with MSE and SSIM loss, respectively. (d) is the FusionNet with adversarial and content loss. (c) shows the legend.}
	\label{fig:network}
\end{figure*}

	To this end, we provide explicit guidance during model optimization to lead the domain shift path  across several identified confident neighborhoods
	, resulting in the proposed framework shown in \fref{fig:network}.
	More specifically, instead of only imposing the loss function on the model output, we introduce multi-scale estimation, multi-branch diversity, and adversarial loss inside the model, thereby  pulling the interim outputs to specific regions then
	 merging them in the target domain; this yields an indirect but more accurate mapping. The contributions of this paper include:

	\begin{compactitem}
	\item By introducing loss functions inside a CNN model, we propose the framework of indirect domain shift (IDS) for image dehazing, which aggregates powerful expressibility of different architectures, i.e., multi-scale, multi-branch, and generator for lifting  degraded images to the target domain indirectly.
	\item We provide theoretic justifications for IDS and show that it provides valuable guidance for network construction.
	    \begin{compactitem}
	    	\item 	A multi-scale module takes the advantage of coarse-fine network to maintain global-local consistency.
	    	\item 	A multi-branch architecture is adopted to enable precise inference of local details by providing diverse confident neighborhoods.
	    	\item 	A FusionNet further improves the perceptual quality by informed `imagination', rather than blindly pursuing a higher PSNR, as the multi-scale multi-branch structure has shifted degraded images close enough to the corresponding ground truth in terms of objective image quality metrics.
	    \end{compactitem}
	\item 	It is demonstrated that IDS leads to remarkable performance improvements compared with the state-of-the-art algorithms.

	\end{compactitem}

\section{Related Works}

  Image dehazing, which aims to recover a haze-free image from its hazy version, is a highly ill-posed restoration problem.
 The haze effect is often approximated using the atmospheric scattering model ~\cite{narasimhan2002vision} given as follows:
	\begin{equation}\label{eq:haze-model}
		\mathbf{I}(x) = \mathbf{J}(x) t(x) + \mathbf{A}(1 - t(x)),
	\end{equation}
  where $\mathbf{I}(x)$, $\mathbf{J}(x)$, and $\mathbf{A}$ are the observed hazy image,  clear scene radiance, and global atmospheric light, respectively. The scene transmission $ t(x)$ describes the portion of light that is not scattered and reaches the camera. It can be expressed as $t(x) = e^{-\beta d(x)}$, where $\beta$ is the medium extinction coefficient and $d(x)$ is the depth map of pixel $x$.

  Based on this atmospheric scattering model~\cite{narasimhan2002vision}, many strategies have been proposed by taking advantage of various prior knowledge.
  For example, the dark channel prior~\cite{he2011single} assumes that in non-sky patches, at least one color channel has very low intensity. 
  The color attenuation prior~\cite{zhu2015fast} assumes that the image saturation decreases sharply at hazy patches, so that the difference between brightness and saturation can be utilized to estimate the haze concentration. To address the weakness of DCP for the sky region, \cite{r11} proposes to separately deal with the non-sky region and the sky region using dark channel prior and luminance stretching. In \cite{r12},  the authors come up with a new color channel method to remove atmospheric scattering for single image dehazing. The overall algorithm consists of atmospheric light calculation, transmission map estimation, radiance estimation and post enhancement. Furthermore, based on the assumption that a linear relationship exits in the  minimum channel between hazy and haze-free images, a fast linear-transformation-based dehazing algorithm is introduced in \cite{r21}.

  Recently, data-driven approaches to image dehazing have received increasing attention.
  ~\cite{ren2016single} and~\cite{cai2016dehazenet} propose to use CNN for medium transmission estimation, which is further leveraged to recover the haze-free image.
  In~\cite{ren2016single}, a multi-scale deep neural network is proposed to learn a mapping between hazy images and their corresponding transmission maps.
  A densely connected pyramid network is proposed in~\cite{Zhang:2018uq} to jointly estimate the transmission map, atmospheric light, and dehazed images, while  an effective iteration algorithm is developed in~\cite{deepprior}  to learn the haze-relevant priors.
  \cite{pfd} further embeds the atmospheric model into the designing of CNN and proposes a feature dehazing unit to ensure end-to-end trainable.
  However, it is known that the atmospheric scattering model (ASM) is not valid in certain scenarios~\cite{li2015nighttime}, which limits the applicability of the aforementioned dehazing methods. Unlike those ASM-dependent methods, ~\cite{deng2019deep} integrates multiple models to perform haze removal with attention, and ~\cite{liuICCV2019GridDehazeNet} uses a GridNet-based network \cite{fourure2017residual} to directly predict dehazed images via an ASM-agnostic approach. To further improve the performance in ASM-agnostic setting,  \cite{MSBDN}  propose an multi-scale boosted dehazing network (MSBDN) with boosting strategy and back-projection technique. \cite{hong2020distilling} firstly introduces knowledge distillation in solving dehazing problem. It allows dehazing model learn to dehaze from both ground truths and teacher outputs.

  Many methods that have been developed for other image restoration tasks, e.g. deblurring, denoising, are also highly relevant.
  To remove blurring caused by the dynamic scenes, a multi-scale convolutional neural network is proposed in~\cite{Nah:2017bx} to restore sharp images in an end-to-end manner.
  In~\cite{gu2014weighted}, the weighted nuclear norm minimization (WNNM) problem is studied and applied to image denoising by exploiting non-local self-similarity. This work is later extended to handle arbitrary degradation, including blur and missing pixels~\cite{Yair:2018wa}.
  To tackle the long-term dependency problem, the MemNet~\cite{Tai:2017uh} is proposed by introducing a memory block, consisting of a recursive unit and a gate unit, to explicitly mine persistent memory through an adaptive learning process.
  To make the deep networks implementable on limited resources, a new activation unit is proposed~\cite{Kligvasser:2018ht}, which enables the net to capture much more complex features, thus requiring a significantly smaller number of layers in order to reach the same performance. A super-resolution generative adversarial network (SRGAN)  is developed in~\cite{Ledig:2017jw} to recover high-frequency details and produce more natural-looking images.

\section{Formulation for Indirect Domain Shift}

\newcommand{\manifold}[1]{\mathcal{#1}}
\newcommand{\loss}{\ell}
\newcommand{\todo}[1]{{\color{blue}Todo: #1}}

In this section, we provide a theoretical formulation of the image dehazing problem and propose an indirect domain shift method as an effective approach to obtaining an approximation solution.

Denote the prior distribution of clear images of size $m\times n$ by $p_{X}$, which is defined on a low dimensional manifold $\manifold{M}$ in $\mathbb{R}^{3\times m\times n}$. The image degradation mechanism can be modeled as a conditional distribution $p_{X|Y}$, i.e., given the clear image $x$, a distorted image $y$ is generated according to $p_{Y|X}$. Note that $p_{X}$ and $p_{Y|X}$ induce the joint distribution $p_{X,Y}$ as well as the conditional distribution $p_{X|Y}$; in general, both $p_{X}$ and $p_{Y|X}$ need to be learned from the training data.
Image dehazing can be formulated as a maximum a posterior estimation problem:
\begin{equation}
\hat{x}_{\text{map}}=\arg\max_{\hat{x}\in\manifold{M}} p_{X|Y} (\hat{x}|y).\label{eq:map}
\end{equation}
In practice, one often considers the following alternative formulation:
\begin{equation}
\begin{aligned}
\hat{x}_{\loss} &= \min_{\hat{x}\in\mathbb{R}^{3\times m\times n}}\mathbb{E}\left[\loss(X,\hat{x})|Y=y\right]\\
&=  \min_{\hat{x}\in\mathbb{R}^{3\times m\times n}}\int_{\manifold{M}}p_{X|Y}(x|y)\loss(x,\hat{x})\mathrm{d}x,
\end{aligned}\label{eq:alternative}
\end{equation}
where $\ell$ is a loss function.
In general it is expected that both $\hat{x}_{\text{map}}$ and $\hat{x}_{\ell}$ are close to the ground truth. However, there is no guarantee that $\hat{x}_{\ell}$ belongs to $\mathcal{M}$.

We shall describe an IDS method, which leverages multi-scale estimation and multi-branch diversity to obtain an approximate solution of (\ref{eq:alternative}), then lifts it into $\mathcal{M}$ using the adversarial loss
to produce a candidate solution of (\ref{eq:map}). A network that realizes the IDS method is shown in \fref{fig:network}.

\subsection{Multi-scale Estimation}\label{sec:multi-scale}

Note that \eqref{eq:alternative} requires the knowledge of $p_{X|Y}$, which needs to be estimated from the training data, hence we solve the following approximated version of \eqref{eq:alternative}, i.e.,
\begin{equation}
\hat{x}'_{\ell}=\min\limits_{\hat{x}\in\mathbb{R}^{3\times m\times n}}\int_{\mathcal{M}}p'_{X|Y}(x|y)\ell(x,\hat{x})\mathrm{d}x,
\end{equation}
where $p'_{X|Y}$ is an approximation of $p_{X|Y}$ learned from the training data.
To ensure that $\hat{x}'_{\ell}\approx\hat{x}_{\ell}$ (and consequently close to the ground truth), we need $p'_{X|Y}(x|y)\approx p_{X|Y}(x|y)$ for $x\in\mathcal{M}$ (at least for $x$ in a neighborhood of $y$ that contains the ground truth). However, since the difference between the ground truth and the distorted version $y$ is not negligible, this neighborhood could be quite large, rendering a good approximation of $p_{X|Y}(\cdot|y)$ in this neighborhood difficult to obtain.
Indeed, the number of parameters need to specify $p_{X|Y}(\cdot|y)$ in this neighborhood might be comparable or even larger than the available training data, hence a direct approximation can be highly unreliable, especially considering the fact that the approximation is in general done in a suboptimal way.
For this reason, it is sensible to first approximate $p_{\tilde{X}|Y}$ (with $\tilde{x}$ being a low-resolution version of the ground truth), which itself is an approximation of $p_{X|Y}$ and can be specified by a significantly smaller number of parameters (as compared to $p_{X|Y}$).
In this way, we can get a good approximation of $p_{\tilde{X}|Y}$, denoted by $p'_{\tilde{X}|Y}$, and solve the following optimization problem instead:
\begin{equation}
\tilde{x}_{\tilde{\ell}}=\min\limits_{\hat{x}\in\mathbb{R}^{3\times m\times n}}\int_{\mathcal{M}}p'_{\tilde{X}|Y}(x|y)\tilde{\ell}(x,\hat{x})\mathrm{d}x.
\end{equation}
Since $p'_{\tilde{X}|Y}(x|y)$ is a good approximation of $p_{\tilde{X}|Y}(x|y)$, it is expected that $\tilde{x}_{\ell}$ is close to $\tilde{x}$ and consequently not very far away from the ground truth. Now with $\tilde{x}_{\ell}$ at hand, we can further convert \eqref{eq:alternative} to the following problem:
\begin{equation}
\hat{x}_{\ell}=\min\limits_{\hat{x}\in\mathbb{R}^{3\times m\times n}}\int_{\mathcal{N}(\tilde{x}_{\ell})}p_{X|\tilde{X}_{\ell},Y}(x|\tilde{x}_{\ell},y)\ell(x,\hat{x})\mathrm{d}x,\label{eq:multiscale}
\end{equation}
where $\mathcal{N}(\tilde{x}_{\ell})$ is a neighborhood of $\tilde{x}_{\ell}$ that is large enough to cover the ground truth.
It suffices to have a good approximation $p_{X|\tilde{X}_{\ell},Y}(\cdot|\tilde{x}_{\ell},y)$ over $\mathcal{N}(\tilde{x}_{\ell})$.
The above procedure is repeated until the required neighborhood is small enough.

We assume that the smaller the neighborhood becomes, the fewer number of parameters are needed to specify the distribution defined over this neighborhood and consequently the approximation becomes easier.
Multi-scale estimation is introduced to mimic conventional coarse-to-fine optimization methods and has been widely applied in many computer vision tasks ~\cite{eigen2014depth,eigen2015predicting,ren2016single,Nah:2017bx}.

\subsection{Multi-branch Diversity}

The idea underlying multi-branch diversity is similar. Suppose we adopt two branches with different loss functions, denoted by $\ell_1$ and $\ell_2$, respectively, then \eqref{eq:multiscale} becomes
\begin{equation}
\label{eq:multibranch}
\begin{multlined}
\hat{x}_{\ell}=\min\limits_{\hat{x}\in\mathbb{R}^{3\times m\times n}}\int_{\manifold{N}(\tilde{x}_{\ell_1})\cap\manifold{N}(\tilde{x}_{\ell_2})}\\p_{X|\tilde{X}_{\ell_1},\tilde{X}_{\ell_2},Y}(x|\tilde{x}_{\ell_1},\tilde{x}_{\ell,2},y)\ell(x,\hat{x})\mathrm{d}x.
\end{multlined}
\end{equation}
It should be clear that multi-branch diversity further narrows the region over which the distribution needs to be estimated. In our experiments, we choose $\ell_1$ and $\ell_2$ to be mean square error (MSE) and structural similarity index (SSIM) loss, respectively. 
The reason we choose MSE and SSIM as loss functions is that MSE focuses on the pixel-level difference while SSIM pays more attention to the perceptual quality. See \fref{fig:network} (a) and (b) for the architecture of two  multi-scale estimation branches of the proposed IDS network.

\subsection{Adversarial Loss}\label{sec:adversarial-loss}

The role of the adversarial loss $\ell_{ad}$ is to lift $\hat{x}_{\ell}$ into $\mathcal{M}$. Specifically, consider a neural network subject to the weighted loss $\ell+\lambda\ell_{ad}$, which can be interpreted as solve the following problem:
\begin{align}
\hat{x}_{\ell+\lambda\ell_{ad}}=\arg\max\limits_{\hat{x}\in\mathcal{N}(\hat{x}_{\ell},\lambda)}p_{X}(\hat{x}),\label{eq:adversarial}
\end{align}
where $\mathcal{N}(\hat{x}_{\ell},\lambda)$ is a neighborhood of $\hat{x}_{\ell}$. In general, this optimization problem tends to give a reconstruction that falls into $\mathcal{M}$ since $p_X$ is only positive on $\mathcal{M}$. Note that the size of $\mathcal{N}(\hat{x}_{\ell},\lambda)$ depends on $\lambda$.
Specifically,  $\mathcal{N}(\hat{x}_{\ell},\lambda)$ is large when $\lambda$ is large.
In the extreme case of $\lambda\rightarrow\infty$, we have $\hat{x}_{\ell+\lambda\ell_{ad}}\rightarrow\arg\max_{\hat{x}\in\mathcal{M}}p_X(\hat{x})$; while when $\lambda$ is very small,  $\mathcal{N}(\hat{x}_{\ell},\lambda)$ may have no intersection with $\mathcal{M}$, and in this case \eqref{eq:adversarial} reduces to \eqref{eq:alternative}.
In principle it is desirable to choose the smallest $\lambda$ such that $\mathcal{N}(\hat{x}_{\ell},\lambda)$ intersects with $\mathcal{M}$. It is also worth noting that $p_X$ is in general unknown. So one has to solve a modified version of (\ref{eq:adversarial}) with $p_X$ replaced by $p'_X$, which is an approximation of $p_X$ learned from the training data.

The adversarial loss serves an important role of generating texture details in image restoration. One of the reasons for its success in our framework is that, by leveraging
multi-scale estimation and multi-branch diversity, one can already obtain an good estimate
$\hat{x}_{\ell}$ which is in a narrow neighboring region of $\manifold{M}$, and consequently the generator does not need much ``imagination" to produce a natural-looking image.
However, we observe the similar phenomenon reported in \cite{Ledig:2017jw} that adversarial loss is helpful for faithful reproduction, even though the final PSNR metric is slightly lower.
Nevertheless, we introduce the adversarial loss to obtain better perceptual quality but not expect higher PSNR value.
The relevant ablation study can be found in \sref{sec:ablation-adv}.

\begin{figure}[!t]
	\centering
	\includegraphics[width=0.9\linewidth]{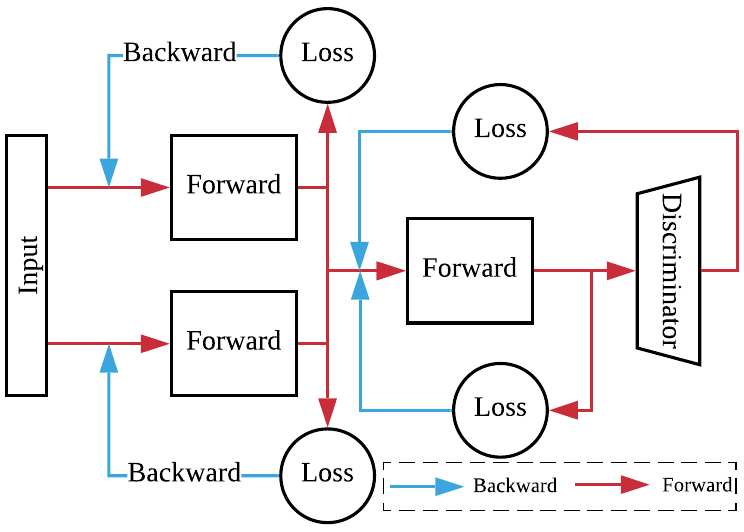}
	\caption{The isolated training of one iteration in hard IDS.}
	\label{fig:training}
\end{figure}

\section{Implementation}

In this section, we provide a detailed implementation of the indirect domain shift (IDS).
We also propose two training schemes, i.e., the hard IDS and soft IDS.

\subsection{Network Architecture} \label{sub:network_architecture}
The proposed IDS network is shown in \fref{fig:network}, which consists of three basic components, i.e., the MSE branch, the MS-SSIM branch, and the FusionNet. The MSE and SSIM branches are built with multi-scale structure to successively map hazy images to their clear counterpart at different resolution levels (as in \eqref{eq:multiscale}); moreover, they are supervised by non-identical loss functions to ensure differentiated outputs.
The FusionNet completes the domain shift process by merging the outputs from the two branches together with the input hazy image into a single clear image (as in \eqref{eq:multibranch}). We train the FusionNet (see \fref{fig:network} (d)) using a content loss defined as the weighted sum of MSE loss and perceptual loss \cite{Perceptual_loss}. The weight is carefully selected by searching from $1.0$, $10^{-1}$, $10^{-2}$, and  $10^{-3}$. We find that our network achieves the best performance when the weight is set to $10^{-2}$. An adversarial loss (see \eqref{eq:adversarial}) is also imposed on the FusionNet to enhance the perceptual quality of the final result.

To be specific, inside each diversity branch, there are three sub-networks, each performing domain shift at a different scale level. The input of the coarse-scale sub-network is obtained from the original hazy image via bi-linear interpolation with a down-sampling factor of 4. Its output is up-sampled with a factor of 2 via pixel shuffle ~\cite{shi2016real},  then fed into the medium-scale sub-network, together with the down-sampled hazy image representation by a factor of 2. The input of the fine-scale sub-network is the concatenation of the original hazy image representation and the up-sampled output of medium-scale sub-network.

It is known that residual networks (ResNets) can facilitate gradient flow while dense networks (DenseNets) help maximize the use of feature layers via concatenation and dense connection. To capitalize on their respective strengths, \cite{Zhang_2018_CVPR} proposes so-called residual dense networks (RDNs), which consist of
contiguous memory blocks, local residual learning blocks and global feature fusion blocks.

In this work, we use
RDNs as the fundamental building components of the proposed IDS network. See
\tref{tab:depth} for detailed specifications. Note that hard IDS and soft IDS adopt the same network structure, but differ in terms of the number of  trainable parameters. Model depth will be detailed in \sref{sec:depth}.

\subsection{Training Scheme}

To handle the coexistence of multiple loss functions,
we propose two back-propagation strategies characterized by different effective ranges of the loss functions. Specifically, we can separately update each module according to the associated loss function or jointly update all modules according to a global loss that aggregates the local ones. This results in the two IDS training schemes, i.e., hard IDS and soft IDS.

\begin{figure}[!t]
  \includegraphics[width=1.0\linewidth]{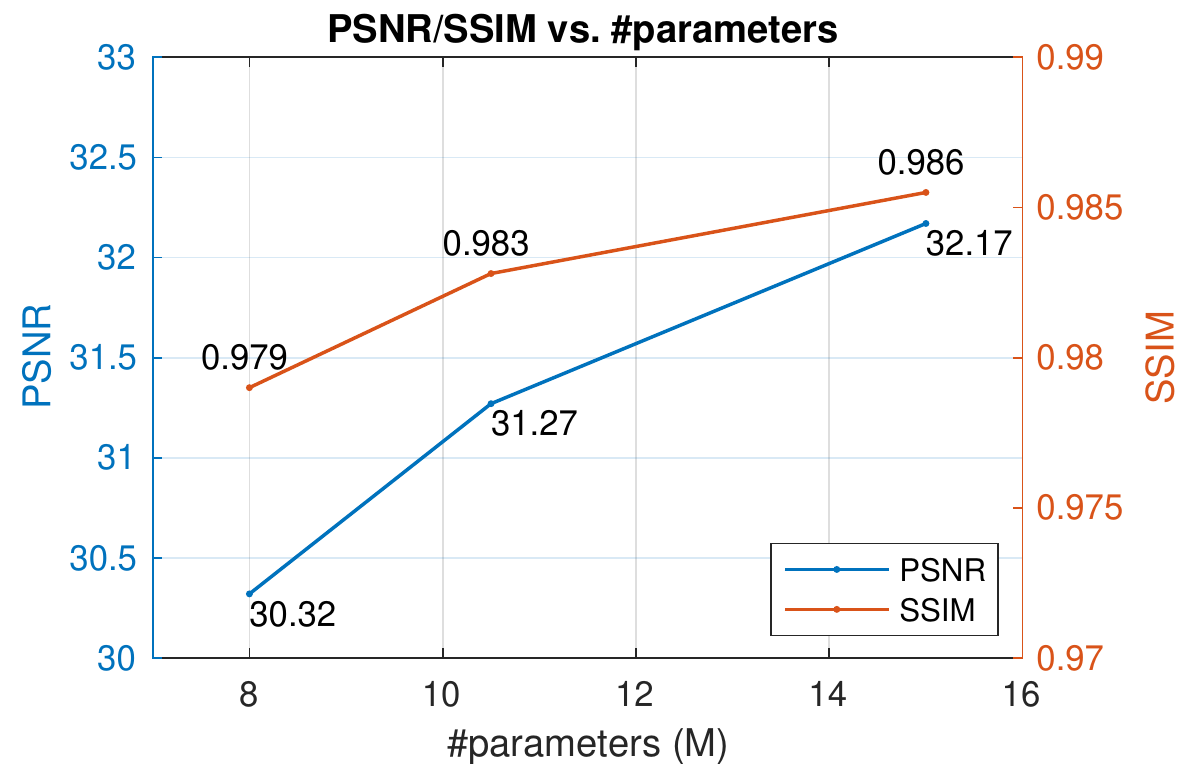}
  \caption{The performance of hard IDS with different parameters.}
  \label{fig:depth}
\end{figure}

\begin{figure*}[!h]
	\centering
	\includegraphics[width=1\linewidth]{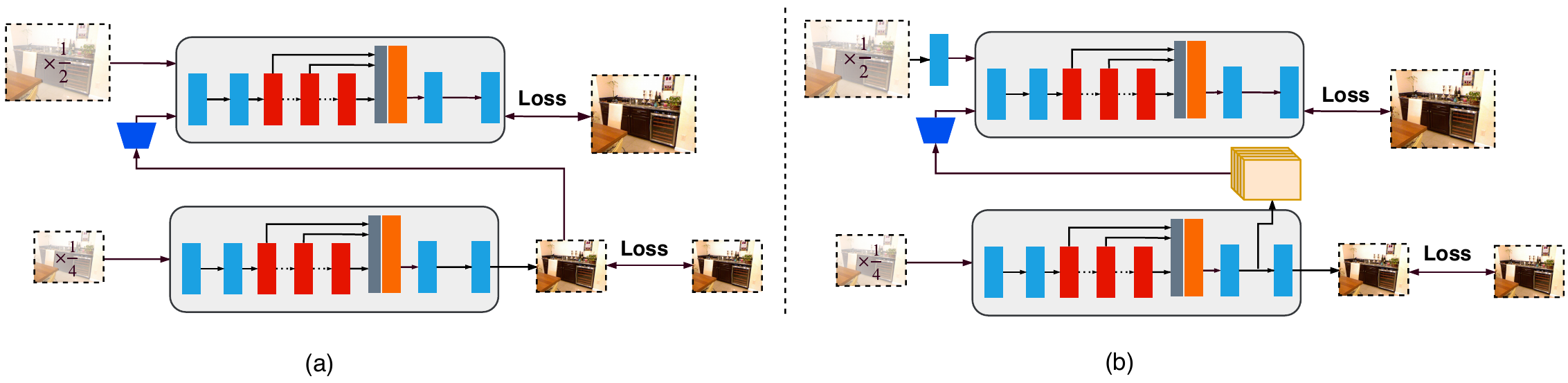}
	\caption{The difference between (a) Hard IDS and (b) Soft IDS.}
	\label{fig:diff}
\end{figure*}

\begin{table}[!t]
  \begin{tabular}{c|l|cccc}\toprule
		\multicolumn{2}{c|}{ \#RDB/\#Conv}    &  Shadow & Medium  &  Deep \\ [3pt] \midrule
		\multirow{3}{*}{Branch} & Coarse & 4/3& 5/3  & 6/3  \\[3pt]
		&  Finer  & 6/4 & 7/4 &  8/4 \\[3pt]
		& Finest &  8/5& 9/5 &  10/5 \\ [3pt]\midrule
		\multicolumn{2}{c|}{FusionNet}  & 10/6 & 12/6 & 15/6  \\[3pt]
		\bottomrule
	\end{tabular}
  \caption{The configuration of the shadow, medium, and deep Hard IDS corresponding to \fref{fig:depth}.}
  \label{tab:depth}
\end{table}
\subsubsection{Hard IDS}
We first present the isolated training strategy for hard IDS shown in \fref{fig:training}. Specifically, each module is supervised independently by the associated loss functions and deliver dehazed images to the next stage after updating their weights.
Note that in this case, the convergence of the entire network does not depend on the convergence of all loss functions, which means that the network performance may become stable before all loss functions are small enough. This is a consequence of direct mapping, since for each mapping step it suffices to enter one of many (almost) equally good confident neighborhoods, resulting in lower computational load.
One advantage of isolated updating is that the gradient vanishing  problem can be  alleviated. Recall that this problem is caused by the emergence of small gradients in the earlier layers of very deep networks during back-propagation. As a comparison, isolated training shortens the back-propagation path, but maintains the depth of forward inference, at the expense of heterogeneous convergence rates of different loss functions.
It is also worth noting that the isolated training strategy closely follows our analytical formulation which dictates how to shift from one domain to another. Therefore, the success of hard IDS can be viewed as a good indication of the correctness of our theoretical framework.

\subsubsection{Soft IDS}

In contrast to hard IDS, here a global loss function obtained by combining all local module losses is used to update network parameters via end-to-end back-propagation. 
Although the local losses are evaluated based on the images output by the respective modules, only the feature map from the penultimate convolutional layer of each module is delivered to the next module. This enables soft IDS to accomplish the desired task largely in the feature space.
The fact that
each module no longer has to re-map the previous module's output images back to the feature space  is helpful for reducing the number of parameters and also making the indirect shifting path `smoother'.
Another advantage of soft IDS is that there is no need to be concerned with the convergence of a specific module as in hard IDS, which facilitates the training process.

In summary, the differences between Hard and Soft IDS are in two main aspects: (1) As in \fref{fig:diff}, Hard IDS and Soft IDS deliver images and features to the next stages, respectively. (2) The Hard IDS adopts isolated training (optimization over modules independently), while Soft IDS computes the summation of all the local module losses and optimizes the entire notwork in an iteration.

\begin{table*}[!t] 
    
    \begin{subtable}{.5\linewidth}
      \centering
        \caption{Hard IDS}
        \begin{tabular}{ccccc}
            \toprule
            Scale & Branch & Adversarial & PSNR & SSIM \\\midrule
            \xmark & \cmark & \cmark & 31.13 & 0.983 \\
            \cmark & \xmark & \cmark & 29.80 & 0.982 \\
            \cmark & \cmark & \xmark & \textbf{31.32} & \textbf{0.986} \\
            \xmark & \xmark & \cmark & 30.55 & 0.975 \\
            \cmark & \cmark & \cmark & \underline{32.17} & \textbf{0.986} \\
            \bottomrule
        \end{tabular}
    \end{subtable}%
    \begin{subtable}{.5\linewidth}
      \centering
        \caption{Soft IDS}
        \begin{tabular}{ccccc}
            \toprule
            Scale & Branch & Adversarial & PSNR & SSIM \\\midrule
            \xmark & \cmark & \cmark & 34.12 & 0.985 \\
            \cmark & \xmark & \cmark & 32.75 & 0.984 \\
            \cmark & \cmark & \xmark & \textbf{34.74} & \underline{0.986} \\
            \xmark & \xmark & \cmark & 33.92 & 0.981 \\
            \cmark & \cmark & \cmark & \textbf{34.74} & \textbf{0.987} \\

            \bottomrule
        \end{tabular}
    \end{subtable}
    \caption{Ablation studies on the SSIM/PSNR performance. The best performance is shown in \textbf{bold}, while second best results are with \underline{underline}.}
    \label{tab:ablation}
\end{table*}

\begin{figure*}[!t]
	\centering
	 \centering
	\begin{subfigure}{.325\textwidth}
		\includegraphics[width=1\linewidth]{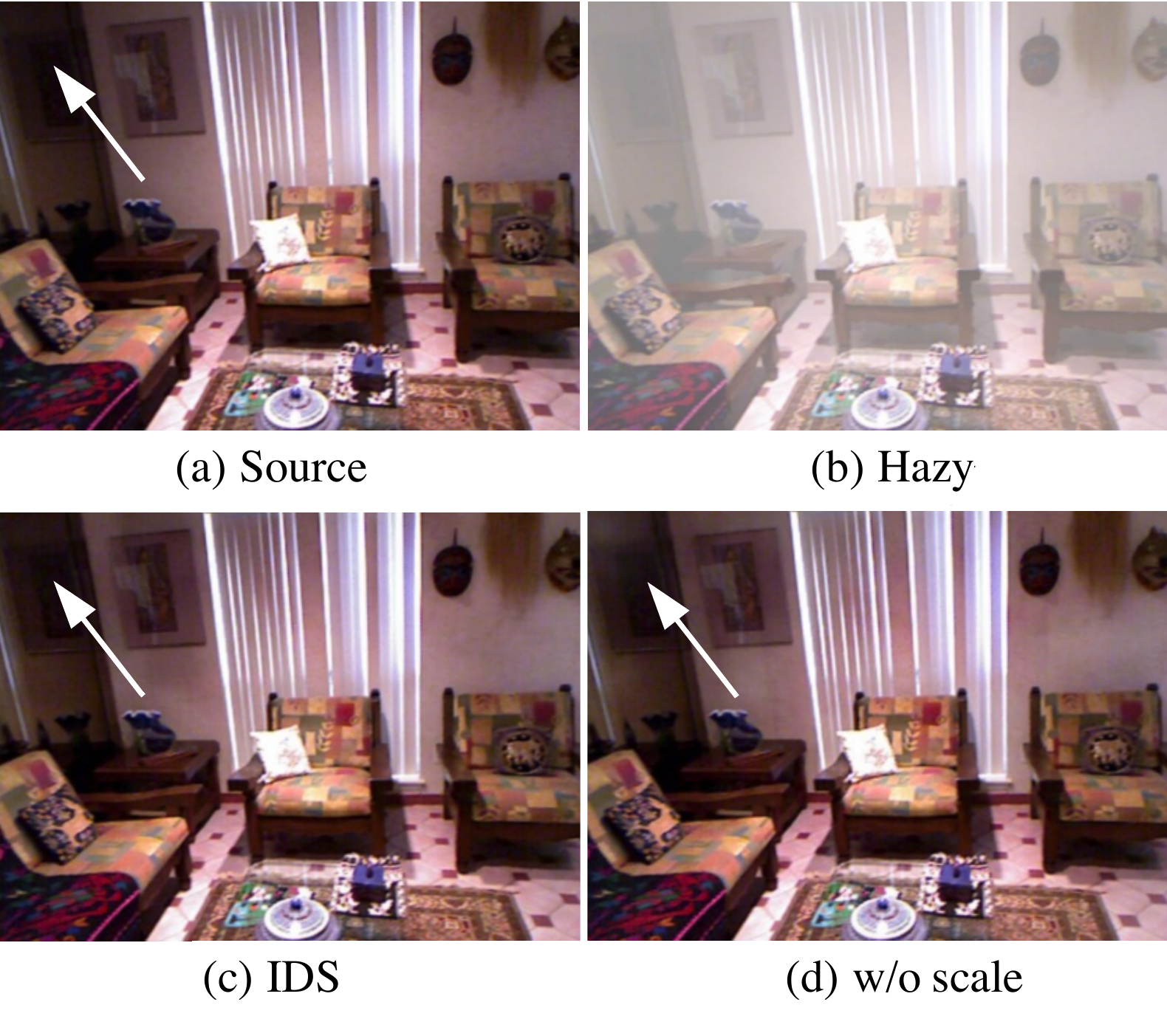}
		\caption{without multi-scale.}
		\label{fig:woscale}
	\end{subfigure}
	\hfill
	\begin{subfigure}{.325\textwidth}
	\includegraphics[width=1\linewidth]{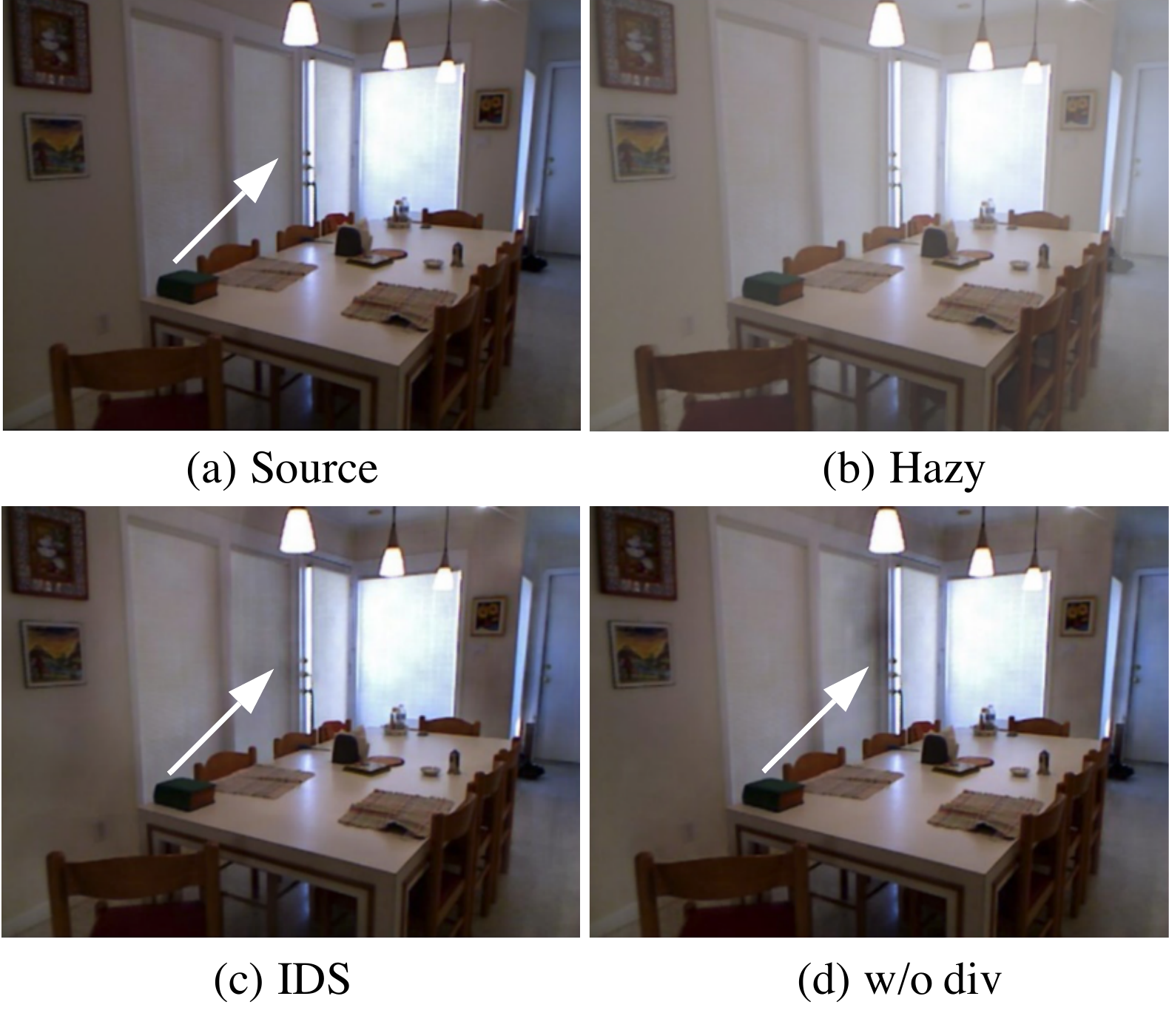}
		\caption{without multi-branch.}
		\label{fig:wodiv}
	\end{subfigure}
	\hfill
	\begin{subfigure}{.325\textwidth}
		\includegraphics[width=1\linewidth]{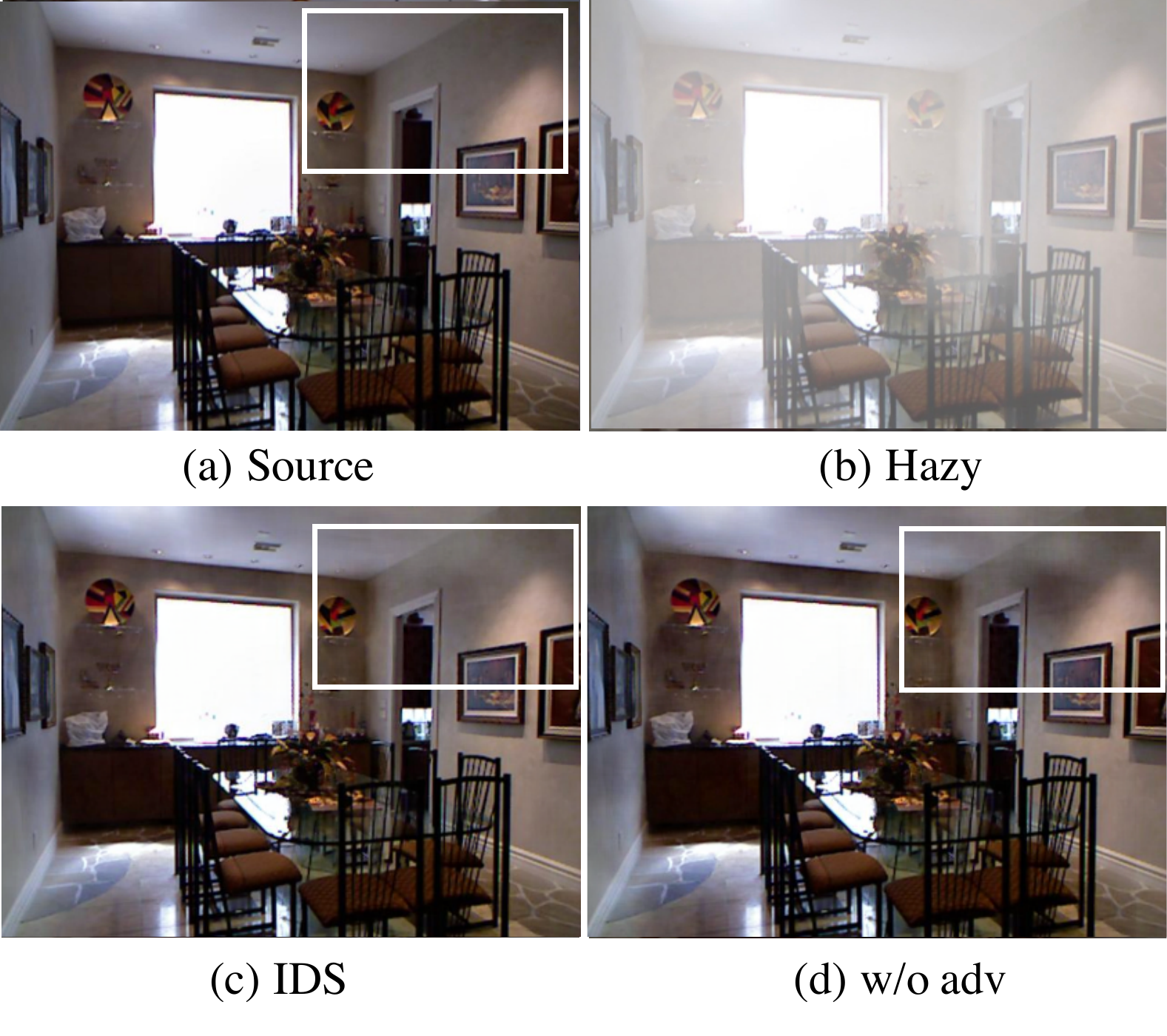}
		\caption{without adversarial loss.}
		\label{fig:woadv}
	\end{subfigure}

	\caption{Some output examples of Hard IDS without multi-scale estimation (w/o scale), without multi-branch diversity (w/o div), only with adversarial loss (o/w adv), and without adversarial loss (w/o adv) in the ablation study, respectively.} \label{fig:abalation}
\end{figure*}

\section{Ablation Study}\label{sec:ablation}

We conduct ablation studies to investigate the respective contributions of multi-scale estimation, multi-branch diversity, and adversarial loss using RESIDE-standard indoor dataset \cite{li2019benchmarking} that will be introduced in detail in \sref{sec:dataset}. To eliminate the influence of other factors, all training configurations are kept the same as that presented in \sref{sec:training}, including the total number of trainable parameters for each network.
More detailed analysis is shown in supplementary.

\subsection{Multi-scale Estimation}

As mentioned in \sref{sec:multi-scale}, a direct mapping can be highly unreliable, since the number of trainable parameters might be comparable or even larger than the available training data. To overcome this problem, a multi-scale network is applied in the first stage of IDS.
Another important property of such coarse-to-fine estimation is the local-global consistency: the coarse-scale network first estimates the holistic structure of the image scene, and then a fine-scale network performs refinement based on both local information and the coarse global estimation.
To further study the influence of such coarse-to-fine structure, we test the performance of IDS framework without multi-scale estimation (w/o scale).

Following the ablation principle, we remove the coarse-scale network and make the fine-scale network deeper to have the same number of parameters.
One output example is presented in \fref{fig:woscale} indicating that hard IDS w/o scale is able to recover the image reasonably well, but with some local inconsistency: the haze at the up-left corner is not removed faithfully. This verifies the above analysis that multi-scale network is able to capture both local and global features. We present the performance on PSNR and SSIM for both hard IDS and soft IDS in \tref{tab:ablation} (a) and (b), respectively.
It can be seen that IDS w/o scale performs worse than IDS (especially in soft IDS), indicating that the local inconsistency has impact on both the quantitative metrics and perceptual quality.

\subsection{Multi-branch Diversity}

Using multi-scale estimation with MSE loss, one can realize domain shift to a certain extent. However, some important information may get lost along the way.
To keep the information diversity, we introduce one more multi-scale branch and employ SSIM loss in this branch.
This strategy enables a more precise inference of local details by providing distinctive confident neighborhoods identified by different branches. To further illustrate its effectiveness of this strategy, we test the performance of IDS without multi-branch diversity (w/o div).

Similarly, we remove the second branch and make the first branch deeper.
One of the examples is presented in \fref{fig:wodiv}, in which the IDS w/o div sometimes delivers erroneous detail inference, since the ``dark area" between the light and the wall clearly should not exist. This is further verified by the overall validation shown in the  \tref{tab:ablation}, in which there is a large performance gap between IDS and IDS w/o div, indicating that it is well worth having two branches.

\subsection{Adversarial Loss}\label{sec:ablation-adv}

The adversarial loss (together with the content loss) is employed at the last stage (i.e., the FusionNet)
of the proposed IDS framework and is served to obtain high visual quality.
The FusionNet takes the estimates from the two branches, in conjunction with the original hazy image, as the input and generates the final output with perceptually satisfactory high-frequency details via proper fusion.
Since the estimates produced by the  two branches are already in the neighboring domains of the target, the generator does not need to rely on pure ``imagination" to create texture details; instead, it could, to a great extent, maintain the perceptual reality rather than blindly pursue a higher PSNR \cite{Ledig:2017jw}.

To prove this, we show that IDS without adversarial loss is able to produce a high PSNR but NOT able to obtain better perceptual quality. Following the ablation principle, we construct IDS
IDS without adversarial loss (w/o adv) by simply removing discriminator.
As can be seen, IDS w/o adv produces a slightly higher PSNR in \fref{fig:woadv} (26.508), but obviously lower perceptual quality than IDS (26.094), as the wall is printed ``darker'' partially to minimize the MSE distance.
This demonstrates the generalization capability of the generator and provides further justifications for the IDS framework.

\begin{figure}[!t]
	\centering
	\includegraphics[width=1.0\linewidth]{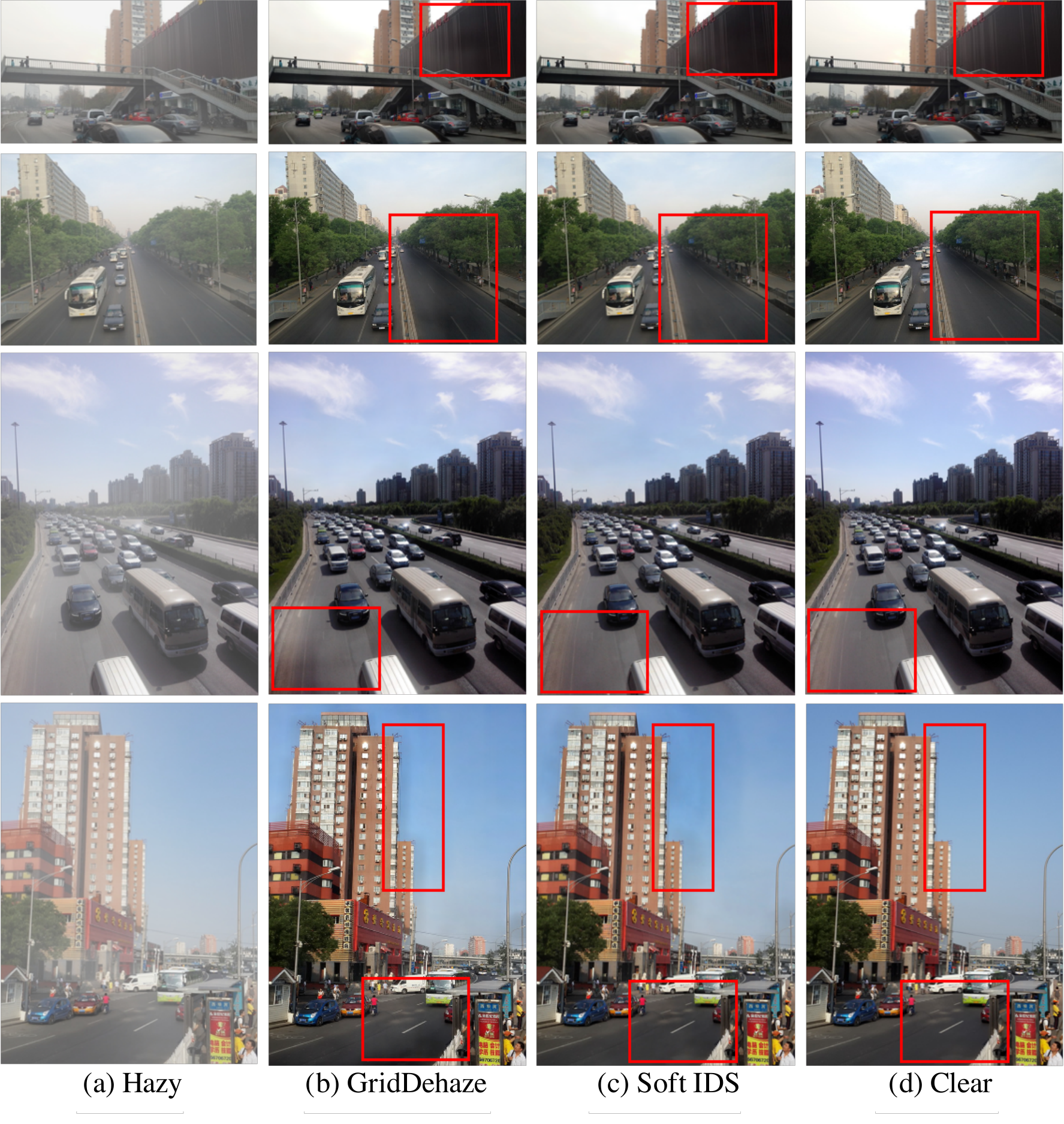}
	\caption{The output examples from SOTS outdoor testing set.}
	\label{fig:outdoor_compare}
\end{figure}

To further prove the necessity of adversarial loss, we compare with GridDehaze \cite{liuICCV2019GridDehazeNet}. GridDehaze \cite{liuICCV2019GridDehazeNet} is a pure CNN based dehazing method without adopting adversarial loss to generate natural distributed outputs.   From \fref{fig:outdoor_compare}, it shows that the generated images from Soft IDS tend to be closer to the ground truth with less inconsistent color gradients on the road, sky, and wall.
This verifies the phenomenon that the adversarial loss is introduced to obtain better perceptual quality but not blindly pursue higher PSNR value.

\subsection{Model Depth} \label{sec:depth}

This section is devoted to investigating the impact of model depth on the performance of our hard IDS method. By adjusting the number of convolutional and dense residual blocks, we construct shadow, medium, and deep models with 8 M, 10.5 M, and 15 M~trainable parameters, respectively. Detailed specifications is shown in \tref{tab:depth}. As expected, the deep model achieves the best overall performance in terms of both PSNR and SSIM. As illustrated in \fref{fig:depth},  both PSNR and SSIM values
improve dramatically as the number of parameters increases, which further verifies the effectiveness of the IDS framework.
It is worth mentioning that
albeit with fewer trainable parameters (around 4.3 M), soft IDS still manages to outperform hard IDS as shown in \tref{tab:quantitative-comp}.

\begin{table*}[!t]
	\caption{The SSIM/PSNR performance of different methods on SOTS-indoor, and SOTS-outdoor. Our proposed methods and improved network with RCAN outperform the others.}
	\begin{center}
	\scalebox{0.85}{
	\begin{tabular}{@{}c|c|c|c|c|c|c|c|c|c|c|c|c @{}}
		\toprule
		Dataset & Metrics & DCP  & DehazeNet  & AOD-Net & GFN  & GridDehaze& PFD& MSBDN & Hard IDS & Soft IDS & RCAN IDS \\ \midrule
		\multirow{2}{*}{Indoor} &PSNR & 16.62 &  21.14 &  19.06 & 22.30  & 32.16 &32.68 & 33.79& 32.17 & \textbf{34.74} & \textcolor{red}{\textbf{35.34}} \\ %\midrule
		&SSIM& 0.8179 &  0.8472 &  0.8504 & 0.880  & 0.9836 &0.9760& 0.9840& 0.9860 & \textbf{0.9869} & \textcolor{red}{\textbf{0.9901}} \\ \midrule
		\multirow{2}{*}{Outdoor}&PSNR & 19.13 & 24.75 &  24.14 & 28.29 & 30.86 &31.17& 31.33&  30.78 & \textbf{31.52} & \textcolor{red}{\textbf{32.73}} \\ %\midrule
		&SSIM& 0.8605 &0.9269  & 0.9198 & 0.9621  & 0.9819 &0.9825&0.9832& 0.9815 & \textbf{0.9832} & \textcolor{red}{\textbf{0.9873}} \\ \bottomrule
	\end{tabular}}
	\end{center}

	\label{tab:quantitative-comp}
\end{table*}

\begin{figure*}[!t]
	\centering
	\includegraphics[width=0.95\linewidth]{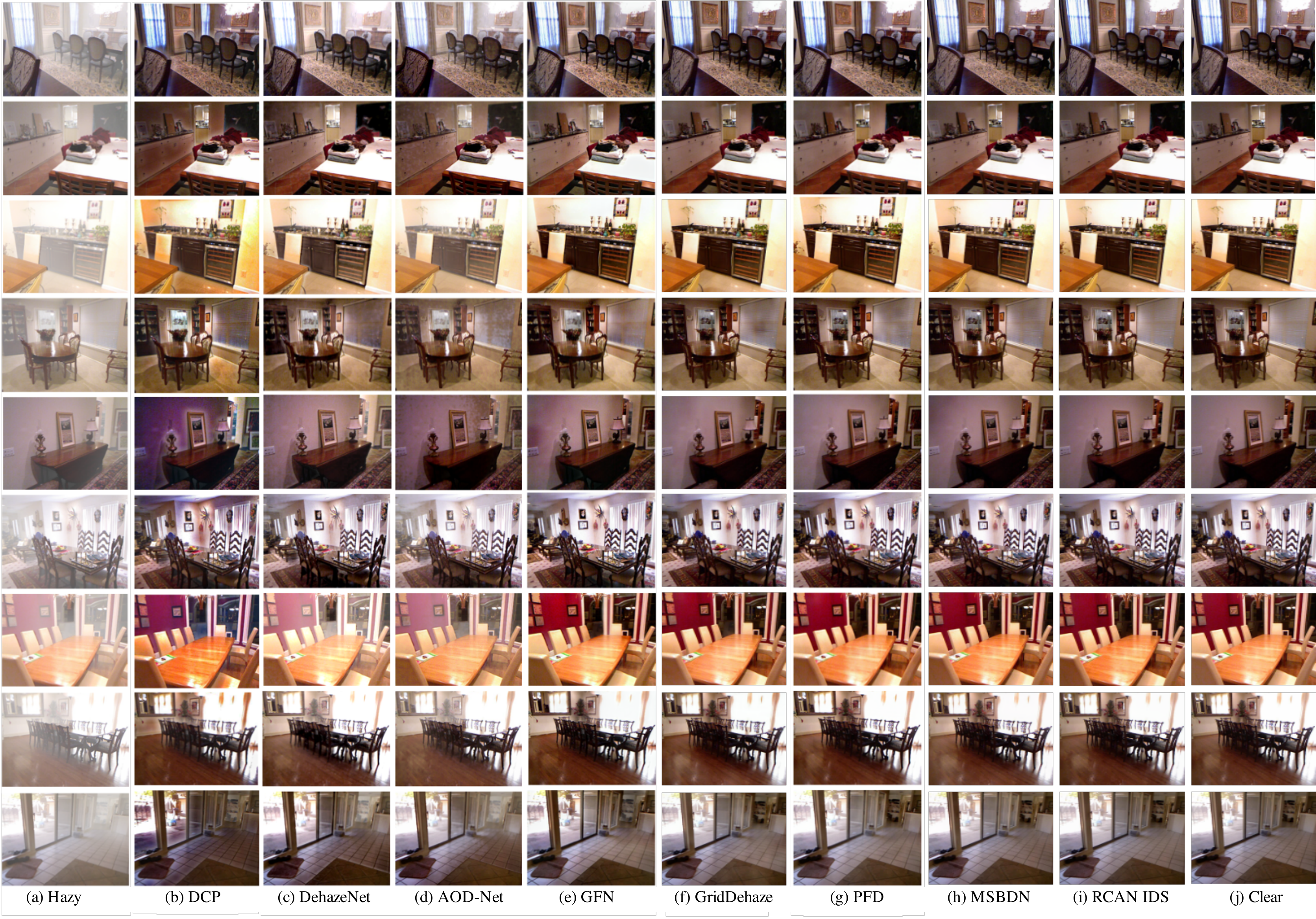}
	\caption{The output examples from SOTS indoor testing set of the SOTA methods.}
	\label{fig:comparison-example}
\end{figure*}

\section{Experiments}

In this section, we further compare the proposed IDS network with several state-of-the-art dehazing algorithms, including dark channel prior (DCP) ~\cite{he2011single}, DehazeNet ~\cite{cai2016dehazenet}, AOD-Net ~\cite{AOD-Net}, gated fusion network (GFN) ~\cite{GFN}, GridDehazeNet (GridDehaze) ~\cite{liuICCV2019GridDehazeNet}, PFD \cite{pfd} and MSBDN \cite{MSBDN}. For a fair comparison, all these algorithms are evaluated on both synthetic and realistic datasets in terms of visual effect and quantitative accuracy.
We adopt the peak signal to noise ratio (PSNR) \cite{zhao2017loss} and the structural similarity index (SSIM) \cite{wang2004image} for evaluation.

\subsection{Benchmark Dataset}\label{sec:dataset}

For training and testing purposes, we use the RESIDE-standard dataset \cite{li2019benchmarking}, which is a benchmark for single image dehazing.
The indoor training set (ITS) of RESIDE-standard contains 13990 synthetic hazy indoor images (together with haze-free counterparts). These synthetic images are generated using NYU2 ~\cite{NYU-V2} and Middlebury stereo ~\cite{middlebury} with the medium extinction coefficient $\beta$ chosen uniformly from $(0.6, 1.8)$ and the global atmospheric light $\mathbf{A}$ chosen uniformly from $(0.7, 1.0)$. The outdoor training set (OTS) of RESIDE-standard contains 296695 hazy images generated from 8477 clear counterparts with $\beta$ chosen uniformly from $(0.04, 0.2)$ and $\mathbf{A}$ chosen uniformly from $(0.8, 1.0)$. The testing set (SOTS) of RESIDE-standard contains 500 synthetic hazy indoor/outdoor images (together with haze-free counterparts). We also perform comparisons using the real-world hazy image dataset in ~\cite{fattal} to show the perceptual difference.

\begin{figure*}[!t]
	\centering
	\includegraphics[width=0.98\linewidth]{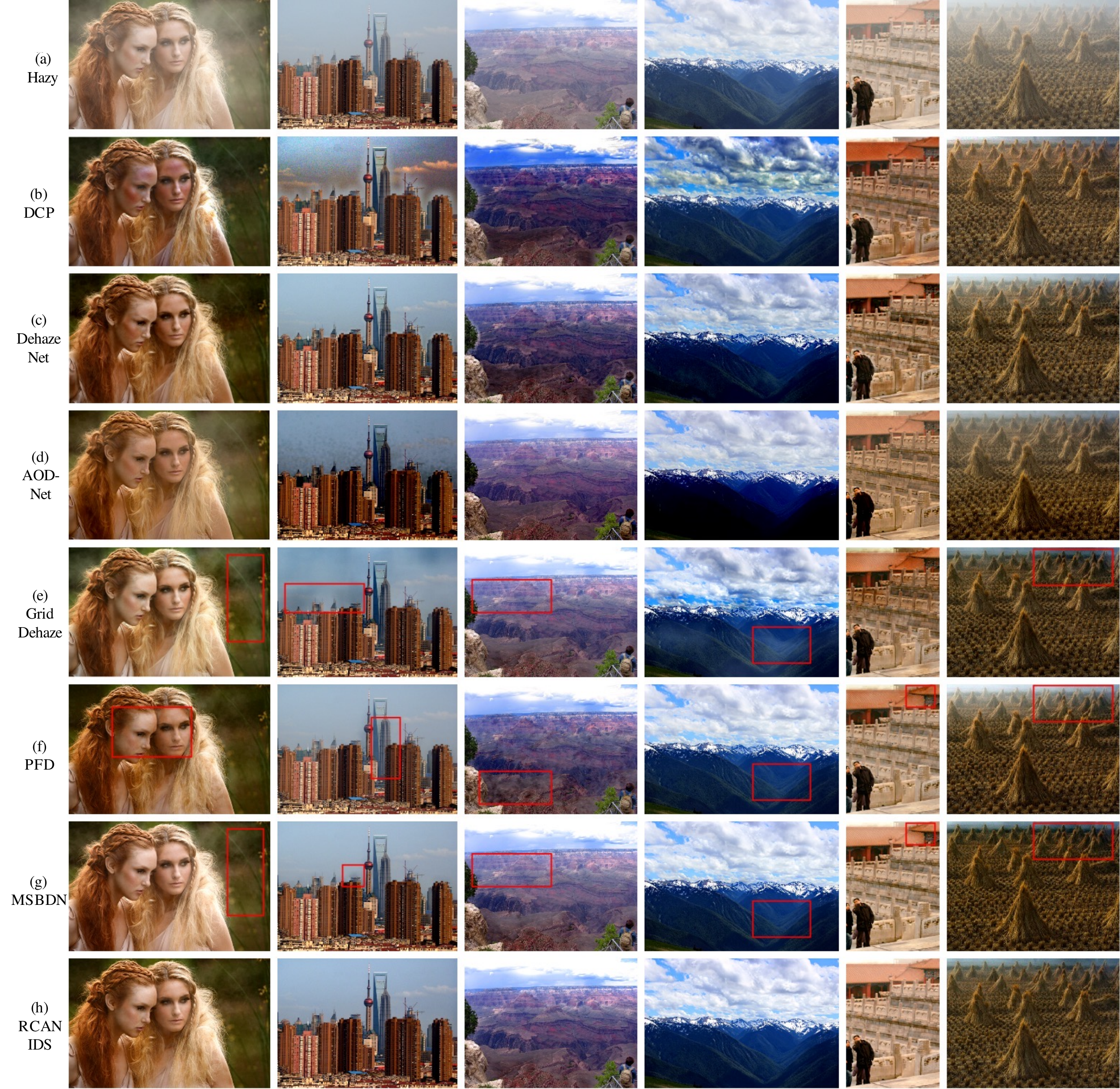}
	\caption{The output examples from real-world images in Fattal et.al.~\cite{fattal} to compare with SOTA DNN based methods.}
	\label{fig:real-world}
\end{figure*}

\subsection{Training Details}\label{sec:training}

Our algorithm is implemented using the PyTorch library \cite{paszke2017automatic} and all tests are conducted on the same GPU of Nvidia Titan Xp.
We train the network with the following configuration: the Adam optimizer \cite{kingma2014adam} is applied with $\beta_1 = 0.9$ and $\beta_2 = 0.999$, where a mini-batch size of 10, a patch size of $180\times180$, an initial learning rate of $10^{-4}$ are adopted.
For hard IDS, the learning rate decays with a multiplicative factor of $0.5$ every 120 epochs for a total of 700  epochs, while soft IDS is trained for 100 epochs with the learning rate reduced by half on the 60th, the 80th, and the 90th epochs. 
Besides,  horizontal/vertical random flipping is applied for data augmentation. It is worth mentioning that after random flipping of both input and target images, the training data are still paired. Therefore, such an augmentation strategy is not harmful to supervised training but help expand the size of training data.

%with learning rate decayed on 50, 70, 90 respectively with factor $0.5$.

\subsection{RCAN as Substitute}\label{sec:substitute}
The proposed IDS framework is generic in nature and admits many different concrete implementations. In this work, we have focused on a particular implementation with RDNs as fundamental building blocks. However, this is by no means the best possible one. Indeed, the performance of our IDS network can be further improved by adopting more powerful substitutes of RDNs. To demonstrate this, we replace RDNs in soft IDS by residual channel attention networks (RCANs) \cite{rcan} with the same number of trainable parameters.  We further illustrate the effectiveness of adopting RCANs as substitute in the following experimental results.

\begin{table*}[!t]\small
	\caption{Average per-image ($620\times460$) runtime (second) on SOTS-indoor.}
	\begin{center}
	\begin{tabular}{@{}c|c|c|c|c|c|c|c|c@{}}
		\toprule
		\multicolumn{9}{c}{DNN Based (GPU Running Time)} \\ \midrule
		DehazeNet & AOD-Net & GFN & GridDehaze & PFD& MSDBN &Hard IDS & Soft IDS & RCAN IDS \\ \midrule
		0.190s  & 0.004s & 0.011s & 0.22s &0.103&0.088& 0.048s & 0.035s & 0.041s  \\ \bottomrule
	\end{tabular}
	\end{center}
	\label{tab:running_time}
\end{table*}

\begin{figure}[!t]
	\centering
	 \centering
	 \begin{subfigure}{1\textwidth}
    	\includegraphics[width=1.0\linewidth]{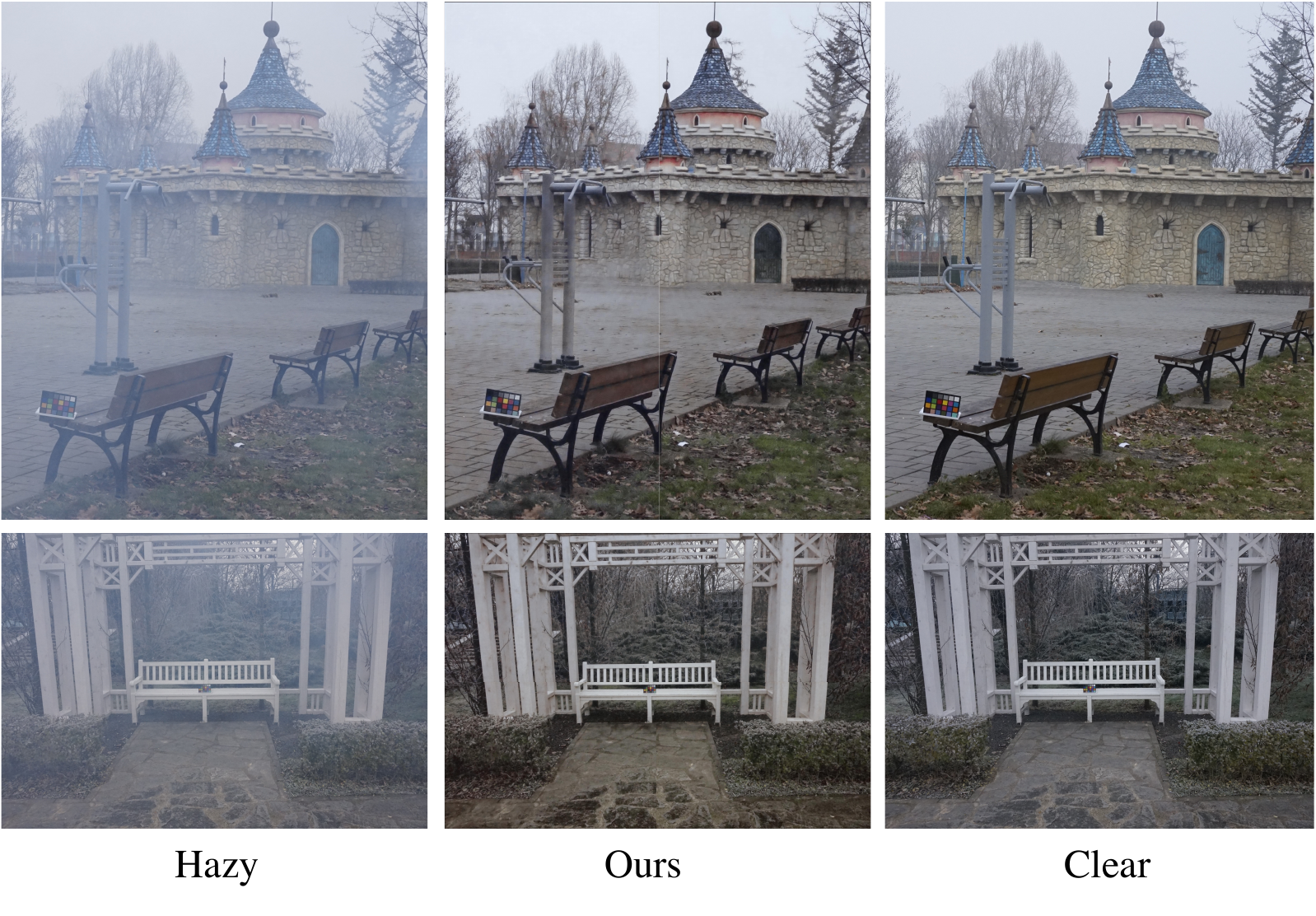}
    	\caption{The output samples from O-Haze testing set.}
    	\label{fig:ohaze}
	\end{subfigure}
	\hfill
	\vspace{\floatsep}
	\begin{subfigure}{1\textwidth}
		\includegraphics[width=1\linewidth]{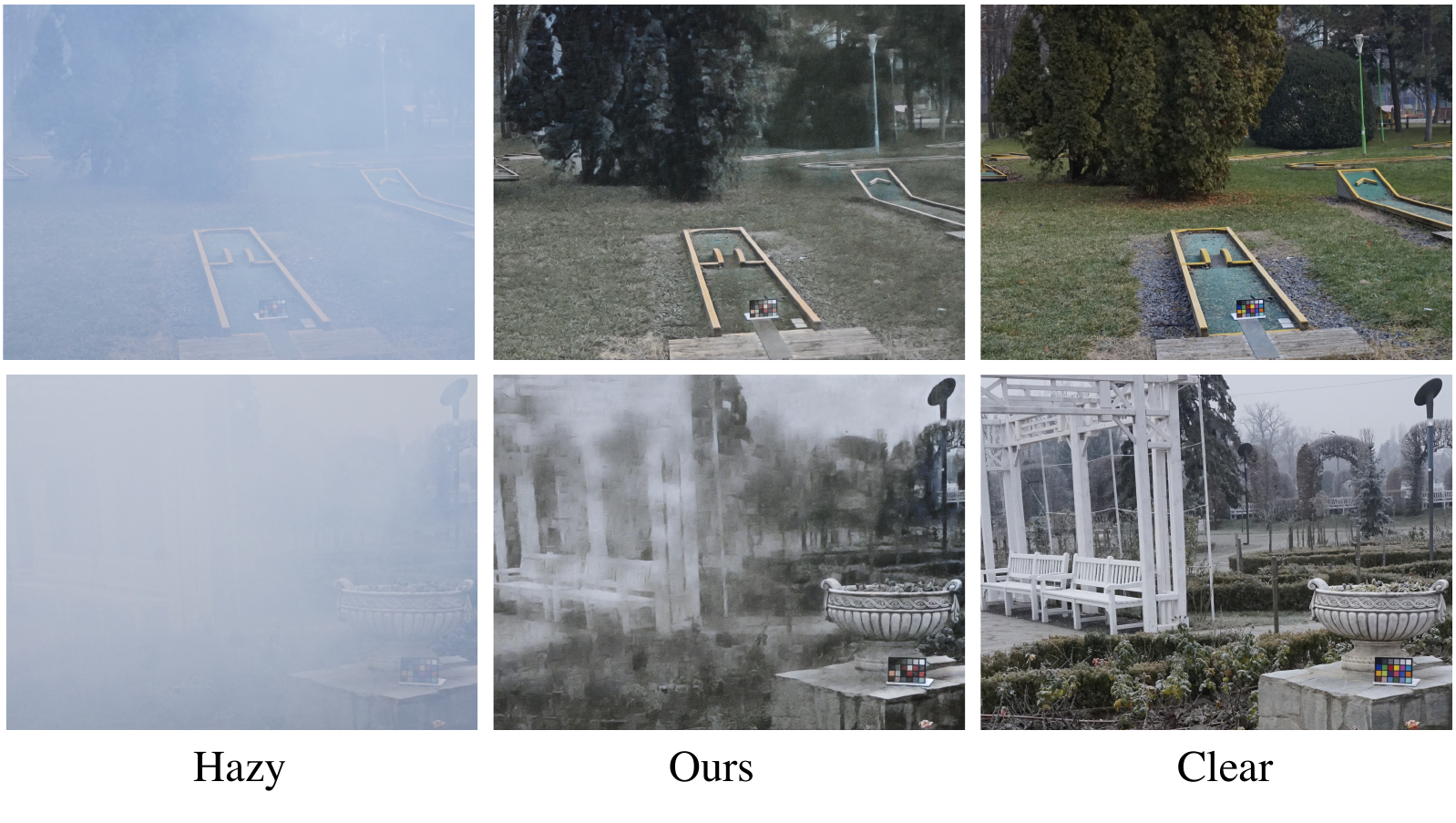}
		\caption{The output samples from Dense-Haze testing set.\\}
		\label{fig:densehaze}
	\end{subfigure}

	\caption{Qualitative evaluation on Dense-Haze \cite{densehaze} and NH-Haze \cite{nhhaze} dataset.} \label{fig:densenhhaze}
\end{figure}

\subsection{Evaluation on Benchmark Dataset}

We train our network  from scratch on RESIDE-standard ITS, OTS and validate it on the separated testing dataset SOTS.  The quantitative results and the qualitative results are shown in \tref{tab:quantitative-comp} and \fref{fig:comparison-example}, respectively. Here hard IDS corresponds to the deep model in \tref{tab:depth}, while soft IDS is as described in \sref{sec:depth}. It can be seen from \tref{tab:quantitative-comp} that  soft IDS outperforms the other methods under comparison in terms of PSNR and SSIM. In particular, the PSNR achieved by  soft IDS reaches 34.74 on SOTS indoor dataset. Moreover, with the boost from RCANs substitute, RCAN IDS outperforms the others by a large margin.

As for visual quality, prior-based methods ~\cite{he2011single} overestimate the haze thickness, which results in color distortion  (e.g. the color of the wall turns purple in the fifth row in \fref{fig:comparison-example}). Although some learning-based baseline methods ~\cite{cai2016dehazenet,AOD-Net} avoid the color distortion problem, they tend to deliver unsatisfactory haze removal results for shaded regions. For example, in the seventh row of \fref{fig:comparison-example}, the area behind the arch should be dark; however, the restoration results produced by most baseline methods show light color instead. This is probably because of that the baseline methods fail to correctly estimate the depth information and consequently mislead by the haze effect. GFN generates decent results, and removes the haze in this area reasonably well. A possible explanation is that GFN does not rely on depth estimation for haze removal; it can also be attributed to the multi-scale approach adopted by GFN, which is an important ingredient of the IDS framework as well. Exploiting the full strength of IDS enables us to obtain better dehazing results.
GridDehaze \cite{liuICCV2019GridDehazeNet}, PFD 
\cite{pfd} and MSBDN \cite{ MSBDN}  are the methods that can produce dehazed images comparable to ours. However, they still generates inconsistent color gradients on the venetian blinds in the fourth row. On the other hand, it can be seen in \fref{fig:comparison-example} that our dehazed images can hardly be distinguished from the ground truth.

\begin{table}[!t]
	\caption{The SSIM/PSNR performance of different methods on O-Haze and Dense-Haze dataset. Our proposed methods  outperform the others.}
	\begin{center}
	\scalebox{0.85}{
	\begin{tabular}{@{}c|c|c|c|c|c @{}}
		\toprule
		\multicolumn{3}{c|}{O-HAZE} &
		\multicolumn{3}{c}{Dense-Haze} \\\midrule
		 Metrics& PSNR & SSIM  &  Metrics & PSNR & SSIM   \\ \midrule
		DCP  &12.92 & 0.505 &  DCP  &  10.85 & 0.404 \\ 
		AOD-Net& 17.69 &  0.616 & AOD-Net& 13.30 & 0.469  \\ 
		GridDehaze& 22.76& 0.721 &  GridDehaze &  14.56 & 0.493  \\ 
		MSBDN &23.28& 0.743 &  MSBDN &  15.18 & 0.509  \\ 
		\textbf{Ours} &\textbf{23.84}& \textbf{0.766} &  \textbf{Ours} &  \textbf{15.78} & \textbf{0.543}  \\ \bottomrule
	\end{tabular}}
	\end{center}
	\label{tab:odensehaze}
\end{table}

\subsection{Evaluation on Real-world Photographs}

We further show the dehazing results on real-world images in \cite{fattal} to illustrate the generalization ability of IDS. In \fref{fig:real-world}, Prior-based method \cite{he2011single}  introduces color distortion and over enhancement on images. 

It is clear that DehazeNet\cite{cai2016dehazenet}, and AOD-Net \cite{AOD-Net} fail to remove haze completely, especially in the last column  where heavy haze can still be seen around the haystack. Moreover, they also tend to over-enhance the images (e.g. the mountains in the fourth column).
 Although GridDehaze \cite{liuICCV2019GridDehazeNet} , PFD \cite{pfd} and MSBDN \cite{MSBDN} work well on the synthetic dataset, its generalization performance on real images is unsatisfactory. The red boxes in \fref{fig:real-world} locate their unsatisfactory regions. Their weaknesses include color distortion, incomplete haze removal and over enhancement.
We also notice that the proposed IDS is able to not only remove haze successfully, regardless whether it is dense or light, but also restore the texture details faithfully, which further proves the effectiveness of our method.

\subsection{Evaluation on Real-world Datasets}

The evaluation is conducted on the O-Haze  \cite{ohaze}, and Dense-Haze \cite{densehaze} datasets. The Two real-world datasets is challenging since they contain limited training images (45 and 55 respectively) and vivid haze patterns. Therefore, the performance on the two dataset can be a good indication to the effectiveness of the proposed methods. The training on the two datasets adopts same strategies as introduced in \sref{sec:training}. For fair comparison, we omit to use pre-trained weights or data augmentations that are not introduced in \sref{sec:training}.
We demonstrate the evaluation quantitatively and qualitatively in \tref{tab:odensehaze} and \fref{fig:densenhhaze}.

\textbf{Results on NTIRE2018 O-Haze.} We evaluate our proposed IDS on O-Haze dataset \cite{ohaze} following the data split in official NTIRE2018-Dehazing challenge \cite{ancuti2018ntire}. It can be observed in \tref{tab:odensehaze} that our IDS outperforms the other methods in terms of PSNR and SSIM.  \fref{fig:ohaze} shows that our approach reconstructs faithful and sharp haze free images with good perceptual quality.

\textbf{Results on NTIRE2019 Dense-Haze.} In contrast to O-Haze that mostly contains light haze, Dense-Haze \cite{densehaze} records images with denser and more homogeneous haze layer. We follows  NTIRE-2019 challenge \cite{ancuti2019ntire} to conduct evaluation.
Qualitative results in \fref{fig:densehaze} demonstrate that even if the background scene is occluded by thick haze, our IDS  is still able to restore these region. In particular, since the second testing sample in \fref{fig:densehaze} is covered by severe haze, the background scene is almost invisible to human eyes. Nevertheless, our IDS surprisingly removes dense haze and reconstructs identifiable details. Quantitative comparisons in \tref{tab:odensehaze} illustrate that our IDS is the top performing method.

\subsection{Runtime}
\tref{tab:running_time} shows runtime comparisons on the SOTS dataset. Our method is ranked the third among DNN-based methods. It is worth mentioning that in our implementation multi-scale estimation is performed branch by branch. A significant reduction in runtime is possible via a parallel  implementation of multi-scale estimation in two branches.

\section{Conclusion}

In this paper, it is shown that the traditional direct mapping methods cannot provide accurate direct mapping for image dehazing.
To solve this problem, an indirect domain shift (IDS) method is proposed by adding explicit loss functions inside a deep CNN model to guide the dehazing process.
Multi-scale estimation, multi-branch diversity, and adversarial loss play important roles in this method as shown by the ablation studies. We also propose two training schemes, which have their respective advantages. 
Specifically, hard IDS is less demanding in terms of computational resources and alleviates the gradient vanishing problem. Besides, hard IDS is designed according to our theoretical formulation and its success provides a strong empirical indication of the correctness of our indirect domain shift mechanism.
On the other hand, soft IDS is easier to implement and in general yields better performance.
We show that IDS achieves remarkable improvements compared with the state-of-the-art on five dehazing datasets.
Despite the success of our method, the visual performance of IDS is not completely satisfactory on Dense-Haze dataset. Since the deep learning methods often require large-scale datasets for training, we believe the performance of our method on Dense-Haze dataset can be further improved by simply acquiring more training samples. From another perspective, one interesting direction for our future work is to enhance the IDS framework to enable good generalization with limited training data.

% \clearpage
% ---- Bibliography ----
%
% BibTeX users should specify bibliography style 'splncs04'.
% References will then be sorted and formatted in the correct style.
%
% \bibliographystyle{splncs04}
% \bibliography{egbib}

% \begin{thebibliography}{1}

% \bibitem{IEEEhowto:kopka}
% H.~Kopka and P.~W. Daly, \emph{A Guide to \LaTeX}, 3rd~ed.\hskip 1em plus
%   0.5em minus 0.4em\relax Harlow, England: Addison-Wesley, 1999.

% \end{thebibliography}

\bibliography{egbib.bib}
\bibliographystyle{plain}

\ifCLASSOPTIONcaptionsoff
\newpage
\fi

\vfill
\end{document}